\documentstyle[epsfig]{mn}
\newif\ifAMStwofonts

\ifoldfss
  \ifCUPmtlplainloaded \else
    \NewTextAlphabet{textbfit} {cmbxti10} {}
    \NewTextAlphabet{textbfss} {cmssbx10} {}
    \NewMathAlphabet{mathbfit} {cmbxti10} {} 
    \NewMathAlphabet{mathbfss} {cmssbx10} {} 
  \fi
  \ifAMStwofonts
    \ifCUPmtlplainloaded \else
      \NewSymbolFont{upmath} {eurm10}
      \NewSymbolFont{AMSa} {msam10}
      \NewMathSymbol{\upi}     {0}{upmath}{19}
      \NewMathSymbol{\umu}     {0}{upmath}{16}
      \NewMathSymbol{\upartial}{0}{upmath}{40}
      \NewMathSymbol{\leqslant}{3}{AMSa}{36}
      \NewMathSymbol{\geqslant}{3}{AMSa}{3E}

    \fi
  \fi
\fi 

\ifnfssone
  \newmathalphabet{\mathit}
  \addtoversion{normal}{\mathit}{cmr}{m}{it}
  \addtoversion{bold}{\mathit}{cmr}{bx}{it}
  \newmathalphabet{\mathbfit} 
  \addtoversion{normal}{\mathbfit}{cmr}{bx}{it}
  \addtoversion{bold}{\mathbfit}{cmr}{bx}{it}
  \newmathalphabet{\mathbfss} 
  \addtoversion{normal}{\mathbfss}{cmss}{bx}{n}
  \addtoversion{bold}{\mathbfss}{cmss}{bx}{n}
  \ifAMStwofonts
    \ifCUPmtlplainloaded \else
      \UseAMStwoboldmath
      \makeatletter
      \new@mathgroup\upmath@group
      \define@mathgroup\mv@normal\upmath@group{eur}{m}{n}
      \define@mathgroup\mv@bold\upmath@group{eur}{b}{n}
      \edef\UPM{\hexnumber\upmath@group}
      \new@mathgroup\amsa@group
      \define@mathgroup\mv@normal\amsa@group{msa}{m}{n}
      \define@mathgroup\mv@bold\amsa@group{msa}{m}{n}
      \edef\AMSa{\hexnumber\amsa@group}
      \makeatother
      \mathchardef\upi="0\UPM19
      \mathchardef\umu="0\UPM16
      \mathchardef\upartial="0\UPM40
      \mathchardef\leqslant="3\AMSa36
      \mathchardef\geqslant="3\AMSa3E
    \fi
  \fi
\fi 

\ifnfsstwo
  \DeclareMathAlphabet{\mathbfit}{OT1}{cmr}{bx}{it}
  \SetMathAlphabet\mathbfit{bold}{OT1}{cmr}{bx}{it}
  \DeclareMathAlphabet{\mathbfss}{OT1}{cmss}{bx}{n}
  \SetMathAlphabet\mathbfss{bold}{OT1}{cmss}{bx}{n}
  \ifAMStwofonts
    \ifCUPmtlplainloaded \else
      \DeclareSymbolFont{UPM}{U}{eur}{m}{n}
      \SetSymbolFont{UPM}{bold}{U}{eur}{b}{n}
      \DeclareSymbolFont{AMSa}{U}{msa}{m}{n}
      \DeclareMathSymbol{\upi}{0}{UPM}{"19}
      \DeclareMathSymbol{\umu}{0}{UPM}{"16}
      \DeclareMathSymbol{\upartial}{0}{UPM}{"40}
      \DeclareMathSymbol{\leqslant}{3}{AMSa}{"36}
      \DeclareMathSymbol{\geqslant}{3}{AMSa}{"3E}
    \fi
  \fi
\fi 

\ifCUPmtlplainloaded \else
  \ifAMStwofonts \else 
    \def\upi{\pi}
    \def\umu{\mu}
    \def\upartial{\partial}
  \fi
\fi

\title[The SW Sex star V348 Puppis]{V348 Puppis: a new SW Sex star in the period gap}
\author[P.~Rodr\'\i guez-Gil et al.]
       {P.~Rodr\'\i guez-Gil,$^1$ I.~G.~Mart\'\i nez-Pais,$^{1,2}$  J.~Casares,$^1$ M.~Villada,$^3$ and L.~van Zyl$^4$ \\
       $^1$ Instituto de Astrof\'\i sica de Canarias, V\'\i a L\'actea, s/n. La Laguna. E-38200. Santa Cruz de Tenerife. Spain \\
       $^2$ Departamento de Astrof\'\i sica, Universidad de La Laguna, Tenerife, Spain \\
       $^3$ Observatorio Astron\'omico de C\'ordoba, Universidad Nacional de C\'ordoba, C\'ordoba, Argentina \\
       $^4$ Department of Astrophysics. Nuclear \& Astrophysics Laboratory. Keble Road. Oxford OX1 3RH}
\date{Accepted 2001.
      Received 2001}

\pagerange{\pageref{firstpage}--\pageref{lastpage}}
\pubyear{2001}

\begin{document}

\maketitle

\label{firstpage}

\begin{abstract}
We present time-resolved optical spectroscopy and photometry of the nova-like cataclysmic variable V348 Puppis. The system displays the same spectroscopic behaviour as SW Sex stars, so we classify V348 Pup as a new member of the class. V348 Pup is the second SW Sex system (the first is V795 Herculis) which lies in the period gap. The spectra exhibit enhanced He\,{\sc ii} $\lambda$4686 emission, reminiscent of magnetic cataclysmic variables. The study of this emission line gives a primary velocity semi-amplitude of $K_1 \simeq 100$ km s$^{-1}$. We have also derived the system parameters, obtaining: $M_1 \simeq 0.65$ M$_{\sun}$, $M_2 \simeq 0.20$ M$_{\sun}$ ($q \simeq 0.31$), $i \simeq 80\degr$ and $K_2 \simeq 323$ km s$^{-1}$. The spectroscopic behaviour of V348 Pup is very similar to that of V795 Her, with the exception that V348 Pup shows deep eclipses. We have computed the ``0.5-absorption'' spectrum of both systems, obtaining spectra which resemble the absorption spectrum of a B0 V star. We propose that absorption in SW Sex systems can be produced by a vertically extended atmosphere which forms where the gas stream re-impacts the system, either at the accretion disc or at the white dwarf's magnetosphere (assuming a magnetic scenario).
\end{abstract}

\begin{keywords}
accretion, accretion discs -- binaries: close -- binaries: eclipsing -- stars: individual: V348 Pup -- novae, cataclysmic variables
\end{keywords}

\section{Introduction}

Cataclysmic variables (CVs) are interacting binaries in which a late-type, main sequence star (the secondary) fills its Roche lobe transferring matter to the 
primary (a white dwarf) through the inner Lagrangian point. Most CVs exhibit 
outbursts during which their brightness suddenly increases, the exception to this being the nova-like CVs (NLs). NLs are permanently found in outburst because of their high ($\sim 10^{17}$ g s$^{-1}$) mass transfer rates (see Warner 1995 for a complete review on CVs).\par

SW Sex systems are a sub-class of NLs which display complex emission and absorption line behaviour (Thorstensen et al. 1991). This group of CVs originally comprised only eclipsing systems with orbital periods in
the range 3--4 hr. Nevertheless, the sample was later extended to include 
non-eclipsing 
systems and to span a wider range of orbital periods (Casares et al. 1996; Smith, Dhillon \& Marsh 1998; Mart\'\i nez-Pais, Rodr\'\i guez-Gil \& Casares 
1999). SW Sex systems exhibit strong single-peaked Balmer,
He {\sc i} and He {\sc ii} emission lines which, with the exception of
He {\sc ii}, remain largely unobscured during primary
eclipse (orbital phase $\varphi=0$). This suggests the presence of emission above the orbital plane. In addition, the radial velocity curves of Balmer and He {\sc i} lines
show significant phase lags relative to the motion of the compact object. The same 
lines also display absorption components at phase
opposite primary eclipse ($\varphi \approx 0.4$--0.5; Szkody \& Pich\'e 1990). 
During the high state, the radial temperature profile in the inner disc is drastically flatter than expected for steady optically thick discs, i.e. $T \propto R^{-3/4}$ (Rutten, van Paradijs \& Tinbergen 1992). Furthermore, the Doppler maps of Balmer and He {\sc i} show extended emission located at the lower left quadrant [i.e. the $(-V_X,-V_Y)$ region]. Finally, recent discovery of variable circular polarization in LS Peg indicates that the action of magnetic fields on the accretion geometry of SW Sex stars (Rodr\'\i guez-Gil et al. 2001) could be important.\par

V348 Pup (1H 0709--360, Pup 1) is an eclipsing NL, first identified as a CV by Tuohy et al. (1990). Its orbital period is $P_{\mathrm{orb}}=2.44$ h (Tuohy et al. 1990; Baptista et al. 1996; Rolfe, Haswell \& Patterson 2000), and it 
displays strong He {\sc ii} emission. The system was detected by the {\it HEAO 1}, {\it Uhuru} and {\it Ariel 5} X-ray surveys, suggesting that it is persistently bright at X-ray energies (Tuohy et al. 1990). {\it ROSAT} also detected the X-ray counterpart of this CV, but with no evidence for orbital modulation (Rosen et al. 1994). The strong He {\sc ii} emission and the detection of X-rays from the system suggests that V348 Pup could be an Intermediate Polar (IP) CV. V348 Pup also exhibits a persistent modulation in its optical light curve with a period 6 per cent longer than the orbital period (Thomas 1993; Rolfe et al. 2000). This is interpreted as the superhump period, caused by the slow precession of an eccentric accretion disc.\par

\section[]{Observations and data reduction}

\subsection{Spectroscopy}

The spectroscopic observations of V348 Pup were performed on 1999 October 24--26 with the 2.15-m Ritchey-Chr\`etien telescope at the Complejo Astron\'omico el Leoncito (CASLEO) in San Juan, Argentina. A total of 33 spectra were acquired on the REOSC spectrograph, equipped with a 600 lines mm$^{-1}$ grating and a Tek 1024~$\times$~1024 pixel$^2$ CCD detector. The instrumental setup gave an spectral range of $\lambda\lambda$3860--5340 at 3 \AA~spectral resolution. The exposure time was fixed at 600 s and spectra of a Cu--Ar comparison arc lamp were taken regularly after 1-2 spectra of the target to secure an optimal wavelength calibration.\par

All the individual frames were de-biased, flat-fielded and sky-subtracted in the standard way. The spectra were then optimally extracted (Horne 1986). The reduction processes were performed using {\sc iraf}\footnote{{\sc iraf} is distributed by the National Optical Astronomy Observatories} routines, whilst for wavelength calibration and most of subsequent analyses we used the {\sc molly} package. A second-order polynomial was fitted to the arc data, the {\sl rms} being always less than 0.12 \AA. Finally, the spectra were normalized to the continuum and re-binned into an uniform velocity scale. The full data set covers $\sim 2.8$ orbital periods of the system.

\subsection{Photometry}

We obtained $R$-band photometric data during a full orbital cycle. The observations were made on 1999 November 21 with the 1.0-m telescope at the South African Astronomical Observatory (SAAO) and the STE4 1024~$\times$~1024 pixel$^2$ CCD detector. The exposure time was 30 s which, together with the read-out time and overheads, resulted in a time resolution of 45 s. We also performed $V$-band photometry during an entire eclipse of the system. The photometric data were obtained on 2001 April 8 with the 1.0-m Optical Ground Station (OGS) telescope at the Observatorio del Teide in Tenerife, from images taken with a Thomson 1024~$\times$~1024 pixel$^2$ CCD camera. The exposure time was 90 s.\par

After de-biasing and flat-fielding the individual images, the instrumental magnitudes were obtained with the PSF-fitting packages within {\sc iraf}. From the scatter in the comparison star light curves, we estimate that the differential photometry is accurate to $\sim1$ per cent.

\section{Spectroscopic analysis}

V348 Pup does not show significant night-to-night variability (as could be 
observed by inspecting each night averaged spectrum). The full averaged 
spectrum (see Fig.~1) shows the typical CV emission lines, namely,  Balmer 
(H$\beta$--H8) and He\,{\sc i} ($\lambda$5015, $\lambda$4922, $\lambda$4471, $\lambda$4026  and other less intense) as well as the high excitation lines He\,{\sc ii} $\lambda$4686, the Bowen blend and C\,{\sc ii} $\lambda$4267. The Ca\,{\sc ii} lines are seen in absorption, with $\lambda$3968 lying in the core of H$\epsilon$. Another absorption feature is visible close to 5200 \AA, which we identify as Fe\,{\sc ii} $\lambda$5169. The presence of this line suggests that the He\,{\sc i} $\lambda$4922 and He\,{\sc i} $\lambda$5015 lines are contaminated by Fe\,{\sc ii} $\lambda$4924 and Fe\,{\sc ii} $\lambda$5018, respectively. Remarkably, the strength of He\,{\sc ii} $\lambda$4686 is larger than H$\beta$. This has also been observed in SW Sex itself during low state (Groot, Rutten \& van Paradijs 2001).

The emission lines show a single-peak profile in the full-orbit averaged spectrum. This is a characteristic of SW Sex systems. To inspect the shape of the line profiles in different parts of the orbit, we constructed averages from the individual spectra taken at the orbital phase ranges 0.35--0.55, 0.75--0.90 and 0.9--1.1 (orbital phases are calculated from the ephemeris given in \S~3.1). These intervals correspond to the expected phase in which the typical absorption component in SW Sex systems becomes stronger, the phases of the hot spot and the phases of eclipse, respectively. The averaged spectra are shown in Fig.~2. In phases 0.35--0.55 the Balmer and He\,{\sc i} profiles are double-peaked, whilst the He\,{\sc ii} $\lambda$4686 and the Bowen blend profiles remain single. In addition, the flux of the Balmer and He\,{\sc i} lines decrease with respect to He\,{\sc ii} $\lambda$4686 (see e.g. the drastic fading of He\,{\sc i} $\lambda$4471). This effect seems to be stronger as we move to higher line excitation levels (i.e. the flux decrease is larger in H$\epsilon$ than in H$\beta$). Double peaks and line fading are caused by the presence of an absorption component reaching maximum strength in this phase range ($\varphi=0.35$--0.55), and constitute a defining feature of SW Sex systems.

The phase 0.75--0.90 average shows the typical emission spectrum, but now He\,{\sc ii} $\lambda$4686 is slightly weaker in flux than H$\beta$. Given that the flux of the Balmer lines does not change significatively outside the 0.35--0.55 phase interval, this fading of He\,{\sc ii} $\lambda$4686 is probably caused by an absorption component, as can be seen in the trailed spectra we show in \S~3.1. During eclipse (phases 0.9-1.1), the averaged spectrum does not differ too much from the previous one. Taking into account that V348 Pup exhibits a deep eclipse, this may indicate that the continuum is more deeply eclipsed than the lines, suggesting line emission originating in material above the plane of the disc. But only by inspecting the line equivalent width (EW) behaviour, we can check whether this affirmation is correct (see \S~3.1.2).

In Table~1, we present some line parameters as measured in the averaged spectrum. The centres and FWHMs were obtained by fitting individual gaussian functions to the line profiles. Uncertainties are rather large because of the low signal-to-noise ratio of the individual spectra.

\subsection{Orbital variability}

Orbital phases have been calculated using the photometric ephemeris given by Rolfe et al. (2000):$${\rm T_0(HJD)= 2448591.667969(85)+0.101838931(14) E},$$

\noindent
were T$_0$ is the time of mid-eclipse.

\subsubsection[]{Radial velocity curves}
We have searched for orbital modulation in the emission lines by extracting velocities through convolution with gaussian templates. The convolution was performed on a window suitably selected for not including any contaminating feature. The FWHM of 
each template was the same than that of the corresponding line, as measured in the averaged spectrum (see Table~1). The radial velocity curves of H$\beta$ and He\,{\sc ii} $\lambda$4686 are shown in Fig.~3. We then fitted the velocity curve of each line with a sinusoidal function of the form:$$V_r=\gamma-K \sin \left[ 2\pi \left( \varphi-\varphi_0 \right) \right].$$

\noindent
In Table~2 we show the fitting parameters for the strongest spectral lines. All the lines are delayed with respect to the motion of the white dwarf by $\varphi_0 \simeq0.1$--0.2. This phase lag is a well known characteristic of SW Sex stars (see e.g. Thorstensen et al. 1991), and indicates that the emission site is at a small angle to the line of centres between the centre of mass and the white dwarf. Interestingly, He\,{\sc ii} $\lambda$4686 is also delayed ($\varphi_0=0.15$), indicating that the bulk of this emission is located close to the region where Balmer emission forms, although the velocity semi-amplitude is smaller.

\subsubsection[]{Equivalent Widths}

In Fig.~4 we present the variation of the EW of H$\beta$, H$\gamma$, He\,{\sc ii} $\lambda$4686 and He\,{\sc i} $\lambda$4471 versus orbital phase. These lines share a very similar behaviour: the EW remains almost constant along phases outside eclipse, undergoing a sudden increase when it begins. This indicates that, for that phase range, the decrease of the continuum flux is greater than that of the line flux. In other words, the lines are less eclipsed than the continuum, as we already proposed above. This behaviour is characteristic of eclipsing SW Sex systems (see e.g. Dhillon, Jones \& Marsh 1994; Still, Dhillon \& Jones 1995), in which either the lines are not obscured during the eclipse (Balmer lines) or they are less obscured than the continuum (He\,{\sc ii} $\lambda$4686 sometimes). In fact, 
this is one of the observational properties defining this family of objects (see Thorstensen et al.\ 1991). Unfortunately, the spectra are not calibrated in flux, so we can say nothing about whether a given line is more deeply eclipsed than another one.

\subsubsection{Trailed spectra}

In Fig.~5 we show the trailed spectra of H$\beta$, He\,{\sc ii} $\lambda$4686 and He\,{\sc i} $\lambda4471$. All the images were constructed by binning the spectra into 15 phase bins, each of which corresponds to a horizontal line in the images. H$\beta$ is dominated by a single emission S-wave which extends out to $\pm 1000$ km s$^{-1}$. The intensity of the emission exhibits variations through the orbit. The strength of H$\beta$ significantly drops around orbital phase $\varphi \simeq 0.4$--0.5. This is due to a wide absorption component crossing the line from red to blue at this point. The detection of such a component in the Balmer and He\,{\sc i} lines constitutes a key fact on the definition of an SW Sex system. We also show the trailed spectra of He\,{\sc i} $\lambda$4471 in Fig.~5. This line is weaker than H$\beta$, so the trailed spectra is noisier, but the line fading at phase $\varphi = 0.4$--0.5 is clearly seen. The He\,{\sc i} $\lambda$4471 absorption is stronger than the same component in H$\beta$. The He\,{\sc i} $\lambda$4471 emission seems to behave in a similar way as H$\beta$, exhibiting a sigle S-wave with roughly the same velocity semi-amplitude. This suggests that both lines form in regions very close together. The structure of the He\,{\sc ii} $\lambda$ 4686 line shows differences from the previous ones. The line profile seems to be the result of the presence of two emission components modulated with the orbital period: one moving with low $K$-velocity ($\approx 100$ km s$^{-1}$) and another less intense moving like the S-wave of H$\beta$, reaching $\pm 1000$ km s$^{-1}$. Like in SW Sex systems, the absorption component is absent from He\,{\sc ii} $\lambda$ 4686. Instead, the line flux seems to decrease at phase $\varphi \approx 0.8$. Both the absorption component of H$\beta$ and the fading of He\,{\sc ii} $\lambda$ 4686 are not reflected in Fig.~4, due to the large errors of the EWs.

\subsection{Doppler tomography}

We have constructed Doppler tomograms of H$\beta$, He\,{\sc ii} $\lambda$4686, and He\,{\sc i} $\lambda4471$ using the maximum entropy method developed by Marsh \& Horne (1988). The maps are shown in Fig.~6. One of the approximations on which Doppler tomography rests lies in that all the line components should be equally visible at all times. So, we constructed the maps only from the individual spectra outside primary eclipse ($\varphi \ne 0.9$--1.1). Besides, the presence of the absorption component around $\varphi \approx 0.5$ also violates this basic premise. We also constructed Doppler tomograms eliminating the spectra where the absorption component is present, obtaining maps that do not differ very much from those shown in Fig.~6. The three crosses on the Doppler maps represent the centre of mass of the secondary star (upper cross), the centre of mass of the system (middle cross) and the centre of mass of the white dwarf (lower cross). The Roche lobe of the secondary and the predicted gas stream trajectory were calculated assuming the system parameters derived in \S~3.3. The stream is marked in steps of 0.1$R_{\mathrm{L_1}}$, where $R_{\mathrm{L_1}}$ is the distance between the white dwarf and the inner Lagrangian point.

The Doppler map of H$\beta$ does not show emission from either the secondary star, the gas stream or the hot spot. The ring-like structure characteristic of the emission from a Keplerian disc around the compact object is also absent from the tomogram. The bulk of emission is concentrated to the lower left quadrant of the map ($V_x <0$, $V_y<0$), extending out to $V_x=-1000$ km s$^{-1}$ (the maximum blue velocity of the S-wave). This emission pattern is characteristic of Balmer lines in SW Sex stars, like V795 Her (Casares et al. 1996), BT Mon (Smith et al. 1998), PX And (Still et al. 1995) or V1315 Aql (Dhillon, Marsh \& Jones 1991; Hellier 1996), and does not have an easy explanation. It could be the 
result of vertical motion of the emitting regions. This cannot be properly handled by the technique of Doppler tomography (see Casares et al. 1996), which assumes that the emission is confined in the orbital plane. Another interpretation has been recently proposed by Horne (1999), based on a magnetic propeller effect, in which the emission comes from gas expelled off the system. One has to be cautious with this model because, it is not clear how the magnetic field is anchored in the disc and how the disc can supply enough angular momentum for the gas to be expelled out.

He\,{\sc ii} $\lambda$4686 shows a flux distribution centred near the expected position of the white dwarf, in agreement once again with the Doppler tomograms of this line in SW Sex stars. This suggests that the bulk of He\,{\sc ii} $\lambda$4686 emission is located closer to the compact object than the Balmer emission. This makes sense if we take a look to the He\,{\sc ii} $\lambda$4686 trailed spectra in Fig.~5. The emission is dominated by a narrow low-velocity component which probably follows the motion of the white dwarf. We will later use this fact to derive the system parameters. On the other hand, He\,{\sc i} $\lambda4471$ again shows emission mainly in the lower left quadrant of the map and displays a crescent-like shape.

\subsection{System parameters}
The intricate line profiles and large phase shifts observed in most eclipsing NLs, makes it very difficult to accurately determine their system parameters. This is especially true in SW Sex stars which, with the exception of the long period system BT Mon, have orbital periods in the range 2.6--4.2 h. This implies that the luminosity of the secondary star in these systems is too low to detect any photospheric absorption feature in the spectra. So, it is impossible to determine the projected orbital velocity of the secondary ($K_2$).\par
However, the Doppler map of He\,{\sc ii} $\lambda$4686 seems to indicate that the emission originates close to the white dwarf. We will use this fact to measure the radial velocity semi-amplitude of the primary ($K_1$). In a first attempt, we calculated the centroid of the He\,{\sc ii} $\lambda$4686 emission on the Doppler map, obtaining a value of $K_1\simeq128$ km s$^{-1}$. But we saw that the flux distribution on the tomogram is not symmetric, causing the centroid not to coincide with the peak of emission and giving an unreliable value of $K_1$. Then, we extracted the emission profile from the map at $V_x=0$, where the white dwarf is located in velocity space. This is shown in Fig.~7. The emission profile is asymmetric and is dominated by a wide component centred at $V_y\simeq-160$ km s$^{-1}$, which extends up to $\pm$1000 km s$^{-1}$. But the peak of the emission comes from a narrow component centred at smaller velocities (i.e. closer to the white dwarf). The profile is accurately fitted by a double-gaussian, as seen in Fig.~7. This supports the presence of two emission components in He\,{\sc ii} $\lambda$4686, as it was previously suggested when discussing the trailed spectra of this line. The double-gaussian fit gives a value of $K_1=100\pm6$ km s$^{-1}$.\par
Now, we will estimate the orbital inclination ($i$). We have obtained $V$-band photometry of V348 Pup to determine the eclipse depth. The light curve is shown in Fig~8. The depth of the eclipse in our $V$-band light curve is $\Delta V\simeq1.6$ mag. This value agrees with the obtained from the {\sc vsnet} database of 1.7 mag. From the $i-\Delta V$ relation derived by Rodr\'\i guez-Gil et al. (2000) for SW Sex systems we get: $$i=5.7\,\Delta V+70.6\simeq80\degr.$$
\noindent This value is in perfect agreement with the obtained by Rolfe et al. (2000) of $81.1\degr \pm 1.0\degr$. On the other hand, the mass of the secondary ($M_2$) can be estimated from the mass-period relation derived by Smith \& Dhillon (1998):
\begin{equation}
\frac{M_2}{{\mathrm M}_{\sun}}=0.126\,P_{\mathrm{orb}}({\mathrm h})-0.11
\label{eq1}
\end{equation}
where $P_{\mathrm{orb}}({\mathrm h})$ is the orbital period expressed in hours. For V348 Pup, we get $M_2=0.20~{\mathrm M}_{\sun}$. The secondary mass function is:
\begin{equation}
f(M_2)=\frac{P_{\mathrm{orb}}K_1^3}{2\pi G}=M_2\left(\frac{q}{1+q}\right)^2 \sin^3 i
\label{eq2}
\end{equation}
Entering $P_{\mathrm{orb}}$, $K_1$, $M_2$ and $i$ we obtain a mass ratio of $q=0.31$, which exactly coincides with the value derived by Rolfe et al. (2000) from the superhump period excess. This indicates that our estimates of $K_1$, $M_2$ and $i$ are correct, especially $K_1$ (given the strong dependence of the mass function on this parameter). In Table~3 we summarize the probable system parameters of V348 Pup.

\section{$R$-band photometry}

The $R$-band light curve of V348 Pup is shown in Fig.~9. Apart from the deep eclipse the light curve seems to display short timescale variability. The modulation caused by the superhump phenomenon detected by Rolfe et al.\,(2000) is also present. We have fitted a sinusoidal function to the data after masking out the eclipse and the ``pulses''. The resulting period is $P_{\mathrm{sh}}=0.108 \pm 0.006$ days, which coincides with the superhump period derived by Rolfe et al.\,(2000).
 
The length of our database is not enough to perform a consistent period analysis, since the short variability could be due to flickering. Consequently, a large database is then needed to reveal the nature of the oscillation. A periodicity of $\sim 15$ min is suggested by eye, although a detailed period analysis with a long database would be required for confirmation. The finding of coherent of semi-coherent (QPOs) oscillations would strengthen the idea of V348 Pup being an IP.

\section{Discussion}

\subsection{V348 Pup is an SW Sex star}

From the evidences presented along this paper we can conclude that V348 Pup belongs to the family of the SW Sex stars. Actually, it satisfies all the conditions stated for a system to belong to this family (see \S~1; see also Thorstensen et al. 1991, where these conditions are established, and Mart\'\i nez-Pais et al. 1999, where they are revised), namely:
\begin{enumerate}
\item The spectra of V348 Pup display single-peaked Balmer, He\,{\sc i} and He\,{\sc ii} emission lines.
\item The emission lines are less eclipsed than the continuum. Contrary to what is observed in SW sex stars, He\,{\sc ii} $\lambda$4686 also exhibits the same behaviour, suggesting that some of the emission forms in regions above the accretion disc. In fact, we have shown that this line have two emission components: a wider one at higher velocities and another one originating very close to the white dwarf (from which we have calculated $K_1$).
\item Balmer and He\,{\sc i} lines show radial velocity curves which are delayed with respect to the motion of the white dwarf. Again, He\,{\sc ii} $\lambda$4686 is also delayed, supporting that some of the emission forms above the orbital plane, close to the regions where Balmer and He\,{\sc i} lines form.
\item Balmer and He\,{\sc i} display an absorption component crossing the lines from red to blue, which reaches maximum strength at orbital phases 0.35--0.55.
\item Doppler tomography of Balmer and He {\sc i} lines shows extended emission located at the lower left quadrant of the maps, exhibiting the characteristic crescent shape.
\end{enumerate}

\subsection{A comparison with V795 Her}
We have shown that V348 Pup is an SW Sex star. Among all these systems, it has the shortest orbital period, which lies in the period gap. The other SW Sex system in the gap is V795 Her, which has an slightly longer orbital period of 2.6 h. The main difference between them is the orbital inclination. While V348 Pup is an eclipsing (high-inclination) system ($i\simeq80\degr$), V795 Her is non-eclipsing (has lower inclination). From geometrical considerations we can say that the inclination of V795 is $i\la70\degr$. The similarity between the orbital periods makes it very interesting to compare the spectroscopic behaviour of both systems.\par
We have used the results of the spectroscopic study performed by Casares et al. (1996) to derive the inclination of V795 Her. The radial velocity curve of He\,{\sc ii} $\lambda$4686 has a semi-amplitude of 85 km s$^{-1}$, which they consider as the $K_1$-velocity of the system. On the other hand, V795 Her also exhibits superhumps (Patterson \& Skillman 1994). The fractional superhump excess, which only depends on $q$, is $\varepsilon=0.07$. Using the relation (Patterson 1998),
$$\varepsilon=\frac{0.23\,q}{1+0.27\,q}$$
to calculate the mass ratio, we obtain $q_{\mathrm{V795}}=0.33$. This value is almost identical to our estimated mass ratio of V348 Pup ($q_{\mathrm{V348}}=0.31$). We can derive the mass of the secondary from equation (\ref{eq1}), which gives $M_2({\mathrm{V795}})=0.22$ M$_{\sun}$. Entering $K_1$, $q$ and $M_2$ in equation (\ref{eq2}) and solving, we get an orbital inclination of $i\simeq53\degr$ for V795 Her (Casares et al. 1996 estimated $i\simeq56\degr$).\par
We have seen that the system parameters of V348 Pup and V795 Her are almost identical, with the exception of $i$, so the main differences between both systems are very likely to be due to inclination effects. The most remarkable difference is that the He\,{\sc i} lines blueward of $\lambda4471$ in V795 Her are fully in absorption. This may suggest that the absorption component is stronger at lower orbital inclination. This is also observed in the non-eclipsing system LS Peg (see e.g. Mart\'\i nez-Pais et al. 1999). On the other hand, the He\,{\sc ii} $\lambda$4686 line in V348 Pup is much stronger than in V795 Her. This can be due to enhanced He\,{\sc ii} emission located above the disc, as we have already suggested.

Now, we are going to compare the FHWM and EW of the Balmer lines H$\beta$, H$\gamma$ and H$\delta$ in both systems. The FWHMs are a factor $\sim 2$ larger in V348 Pup. This is a clear effect of its higher inclination, because the projected velocity of the disc material is larger. The same effect can be seen in the He\,{\sc i} and He\,{\sc ii} lines. The same inclination effect applies to the EWs. They are larger in V348 Pup because the disc is seen at a higher inclination, so the projected surface is smaller and the continuum emission from the disc lower.

\subsection{The spectrum of the absorption component}
As we described in \S~3.1, the Balmer and He\,{\sc i} emission is affected by an absorption component at phases 0.35--0.55. This is one of the defining features of the SW Sex class of CVs. In a recent study, Groot, Rutten \& van Paradijs (2001) found SW Sex itself in low state. During the phases $0.75 < \varphi <0.85$ the bluer Balmer lines changed from emission to absorption up to H14. The spectrum closely resembles that of a B-type star (earlier than B2). The absorption in SW Sex is also present in phases around 0.5, as we can see in the trailed spectra of Balmer (especially H$\delta$) and He\,{\sc i} lines in Groot's study. Having this in mind, we tried to obtain the spectrum of the absorption component at orbital phases 0.35--0.55 in V348 Pup. We assume that the system produces a constant emission spectrum throughout the orbit and that this spectrum is affected by the absorption component at phases 0.35--0.55. As we can see in Fig.~4, the variations in the EWs are small, if we exclude the phases of eclipse. To obtain the absorption spectrum of this component, we decided to subtract the averaged spectrum of V348 Pup corresponding to the orbital phases outside eclipse and outside the absorption events to the averaged spectrum at the absorption phases. Before the subtraction, we corrected all the spectra for orbital motion, using the semi-amplitude of the radial velocity curve of H$\beta$ and the corresponding phase offset (see Table 2). This analysis has to be taken with extreme caution, since the presence of another emission component in certain orbital phases would prevent the subtraction of being the spectrum of the absorption component.\par 
The spectrum of the absorption component in V348 Pup is shown in Fig.~10. The He\,{\sc ii} $\lambda$4686 and the Bowen Blend remain in emission, indicating that the absorption component is absent from these lines. Little emission also remains in H$\gamma$ and H$\delta$. This is due to the different semi-amplitudes and phase offsets of their radial velocity curves with respect to H$\beta$, when correcting for orbital motion. Taking this into account, the Balmer/He\,{\sc i} flux ratios in the absorption spectrum of V348 Pup are very similar to those in the spectrum of a B0 V star (taken from Jacoby, Hunter \& Christian 1984), as can be seen in Fig.~10. This may be a coincidence, but the fact that we get a spectral type similar to the observed by Groot et al. (2001) is encouraging.\par

If the spectrum we obtained is actually the spectrum of the absorption component, and is similar to that of a B star, the absorption events in SW Sex stars could be produced by an atmosphere. Following the magnetic model proposed by Rodr\'\i guez-Gil et al. (2001), we suggest that the absorption is produced by a vertically extended atmosphere. The gas stream from the secondary hits the disc, and part of the stream, which is thicker than the disc itself, overflows it. Another shock is then produced when the stream meets the magnetosphere of the white dwarf. We suggest that an atmosphere is formed around this shock. The temperature of the photospheres of B0-type stars ranges between $19000~\mathrm{K} < T < 25000~\mathrm{K}$, depending on local gravity. We do not expect the local gravity in the shock region to be comparable to that in a giant's photosphere, so the temperature of the absorbing atmosphere will be on the lower end of this range. This temperature should be high enough to produce absorption of H\,{\sc i} and He\,{\sc i}, but not as high as to produce significant He\,{\sc ii} absorption. Another possibility, following the {\sl disc overflow} model (Hellier \& Robinson 1994; Hellier 1996) is that the white dwarfs in SW Sex systems were non-magnetic, so the gas stream re-impacts the accretion disc in its inner regions. This violent impact would form an extended atmosphere which can be responsible of the absorptions.\par


\section*{Acknowledgments}

The 2.15-m ``Jorge Sahade'' telescope at CASLEO observatory is operated under agreement between the Consejo Nacional de Investigaciones Cient\'\i ficas y T\'ecnicas (CONICET) de la Rep\'ublica Argentina and the National Universities of La Plata, C\'ordoba and San Juan. The 1.0-m OGS telescope is operated on the island of Tenerife by the European Space Agency (ESA) in the Spanish Observatorio del Teide of the Instituto de Astrof\'\i sica de Canarias (IAC). We also acknowledge SAAO for the observing time on the 1.0-m telescope.\par
We are grateful to CASLEO staff for their help and support, especially to Antonio de Franceschi. We thank the anonymous referee for her/his helpful comments, which have improved the quality of this paper. We also thank Tom Marsh for the use of his {\sc molly} package.

\bsp

\begin{table*}
 \centering
 \begin{minipage}{80mm}
  \caption{Line parameters measured in the averaged spectrum}
  \begin{tabular}{@{}lccc@{}}
  Line                      & Centre        & EW    & FWHM          \\
                            & (km~s$^{-1}$) & (\AA)     & (km~s$^{-1}$) \\[10pt]
  H$\beta$                  & $54\pm 75$    & 25.2$\pm$1.9 & 1400$\pm$105  \\
  H$\gamma$                 & $2\pm 140$    & 23.6$\pm$2.8 & 1521$\pm$200  \\ 
  H$\delta$                 & $-3\pm 250$   & 22.0$\pm$3.5 & 1916$\pm$365  \\
  H$\varepsilon$            & $17\pm 350$   & 12.9$\pm$3.5 & 1670$\pm$500  \\
  H8                        & $-20\pm 425$  & 10.6$\pm$4.0 & 1535$\pm$ 615 \\
  He\,{\sc i} $\lambda$5015  & $187\pm 950$  &  3.6$\pm$2.0 & 2120$\pm$1600 \\
  He\,{\sc i} $\lambda$4922  & $132\pm 495$  &  4.7$\pm$1.8 & 1620$\pm$755  \\
  He\,{\sc i} $\lambda$4471  & $1\pm 415$    &  5.7$\pm$2.2 & 1416$\pm$650  \\
  He\,{\sc i} $\lambda$4026  & $-103\pm 740$ &  5.8$\pm$3.4 & 1639$\pm$1120 \\
  He\,{\sc ii} $\lambda$4686 & $42\pm 64$    & 31.6$\pm$2.3 & 1293$\pm$95   \\
  Bowen blend               & --            & 11.0$\pm$1.8 & 1555$\pm$400  \\  
\end{tabular}
\end{minipage}
\end{table*}

\begin{table*}
 \centering
 \begin{minipage}{80mm}
  \caption{Parameters of the radial velocity curves obtained for several lines.}
  \begin{tabular}{@{}lccc@{}}
  Line                      & $\gamma$      & $K$ & $\varphi_0$          \\
                            & (km~s$^{-1}$) & (km~s$^{-1}$) & \\[10pt]
  H$\beta$                  & $13\pm 40$    & 275$\pm$55   & 0.162$\pm$0.031 \\
  H$\gamma$                 & $-35\pm 80$   & 216$\pm$115  & 0.127$\pm$0.081 \\ 
  He\,{\sc i} $\lambda$4471  & $-64\pm 195$  &  276$\pm$295 & 0.106$\pm$0.150 \\
  He\,{\sc ii} $\lambda$4686 & $41\pm 40$    & 190$\pm$54   & 0.151$\pm$0.045 \\
\end{tabular}
\end{minipage}
\end{table*}

\begin{table*}
 \centering
 \begin{minipage}{80mm}
  \caption{System parameters of V348 Pup.}
  \begin{tabular}{@{}ll@{}}
  Orbital period, $P_{\mathrm{orb}}$ & 2.444 h \\
  Primary mass, $M_1$       & 0.65 M$_{\sun}$ \\
  Secondary mass, $M_2$     & 0.20 M$_{\sun}$ \\
  Mass ratio, $q$           & 0.31 \\
  Separation, $a$           & $6.07\cdot10^{10}$ cm \\
  Inclination, $i$          & 80$\degr$ \\ 
  $K_1$                     & 100 km s$^{-1}$ \\
  $K_2$                     & 323 km s$^{-1}$ \\
\end{tabular}
\end{minipage}
\end{table*}

\begin{figure*}
\begin{center}
  \mbox{\epsfig{file=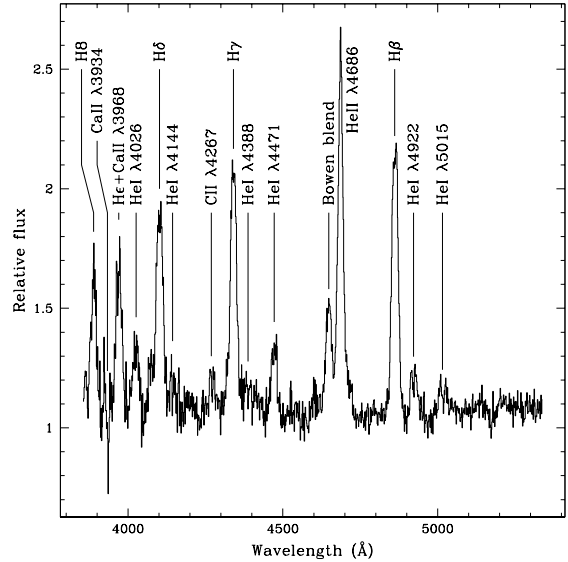,width=8cm}}
  \caption{The averaged spectrum of V348 Pup. All the emission lines are single-peaked, a characteristic of SW Sex systems.}
\end{center}
\end{figure*}

\begin{figure*}
\begin{center}
  \mbox{\epsfig{file=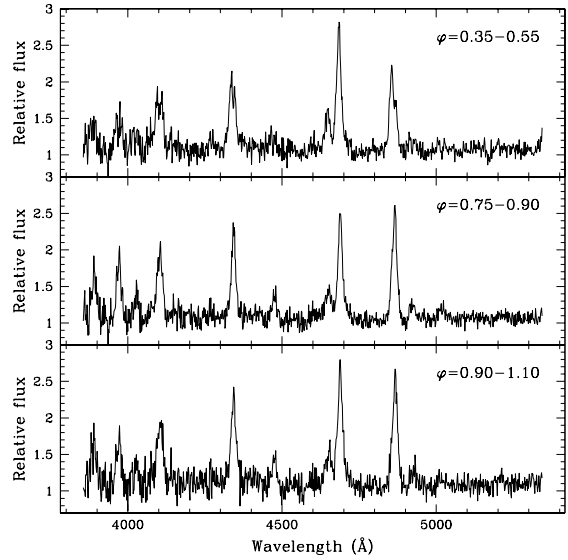,width=8cm}}
  \caption{Averaged spectra of V348 Pup in different phase ranges (see text for details).}
  \end{center}
\end{figure*}

\begin{figure*}
\begin{center}
  \mbox{\epsfig{file=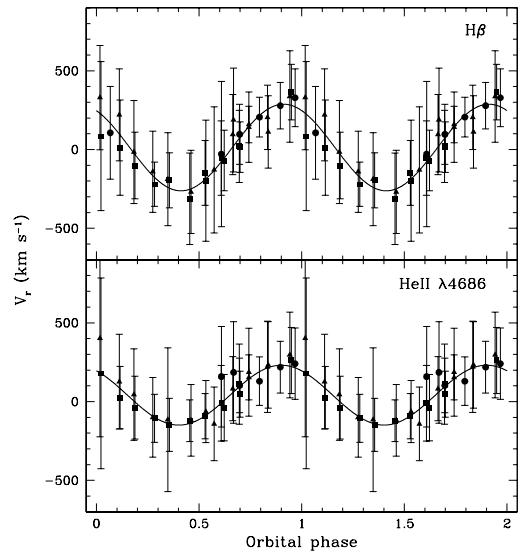,width=8cm}}
  \caption{The radial velocity curves of H$\beta$ ({\it top}) and He\,{\sc ii} 
  $\lambda$4686 ({\it bottom}). Circles correspond to the data obtained during the first 
  observing night, triangles correspond to the second night and squares 
  to the last night. The solid curves are sinusoidal fits to the data (see Table~2). A full orbital cycle is repeated for clarity.}
  \end{center}
\end{figure*}

\begin{figure*}
\begin{center}
  \mbox{\epsfig{file=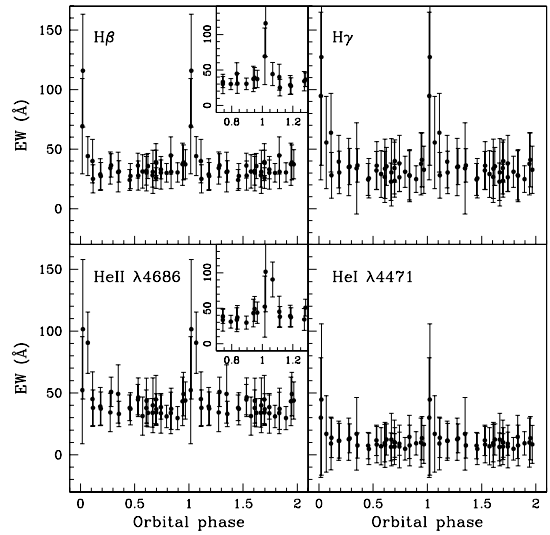,width=8cm}}
  \caption{The variation of the equivalent width (EW) of several emission lines as a function of orbital phase. Note the strong increase in a narrow phase interval around 
  zero phase (i.e. during the eclipse).}
  \end{center}
\end{figure*}

\begin{figure*}
  \mbox{\epsfig{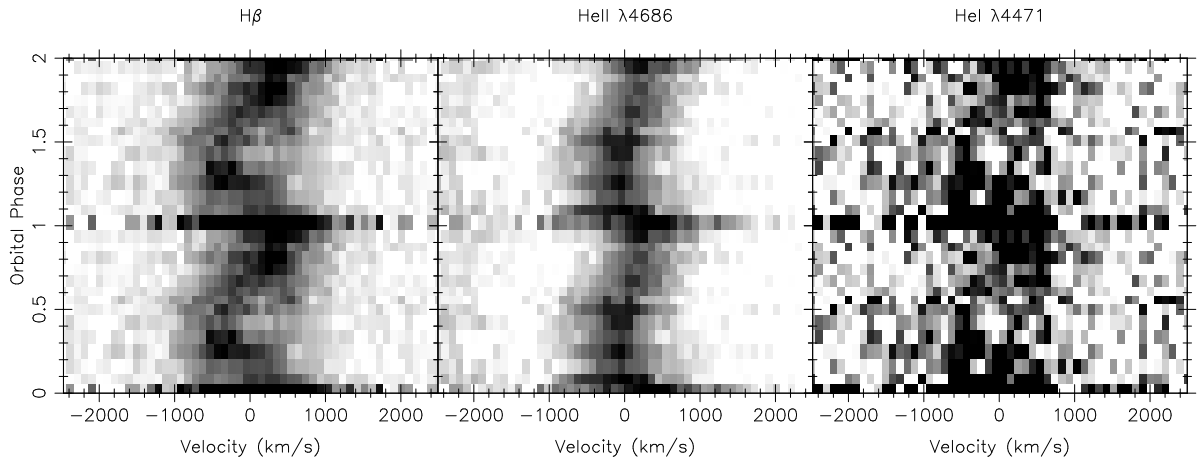}}
  \caption{The trailed spectra of the strongest lines: H$\beta$ and He\,{\sc ii} $\lambda$4686.}
\end{figure*}

\begin{figure*}
  \mbox{\epsfig{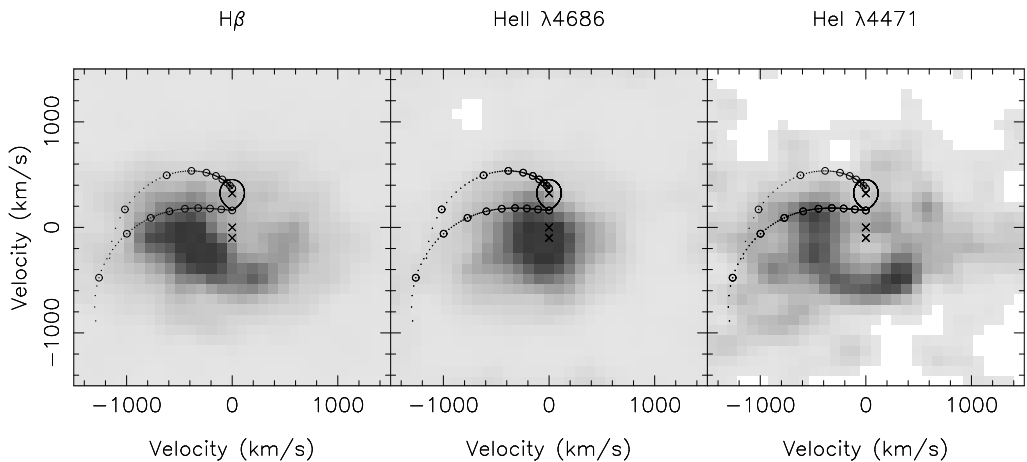}}
  \caption{The Doppler maps of H$\beta$ and He\,{\sc ii} $\lambda$4686.}
\end{figure*}

\begin{figure*}
\begin{center}
  \mbox{\epsfig{file=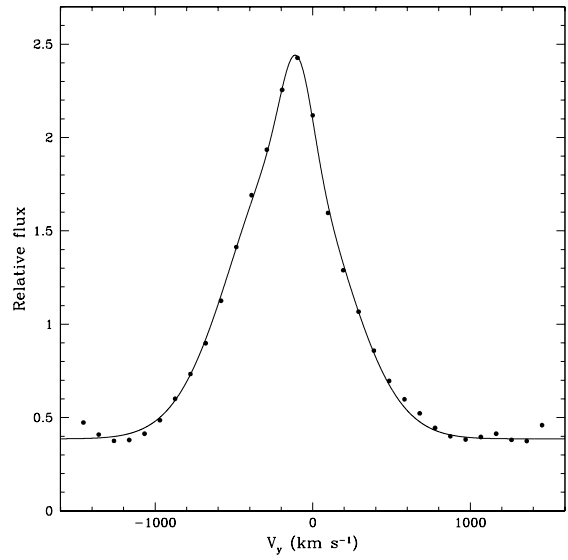,width=8cm}}
  \caption{Emission profile of the He\,{\sc ii} $\lambda$4686 line extracted from the Doppler map at $V_x=0$. The profile is clearly asymmetric and is very well fitted by a double-gaussian function ({\it solid line}), suggesting the presence of two emission components in the line.}
  \end{center}
\end{figure*}

\begin{figure*}
\begin{center}
  \mbox{\epsfig{file=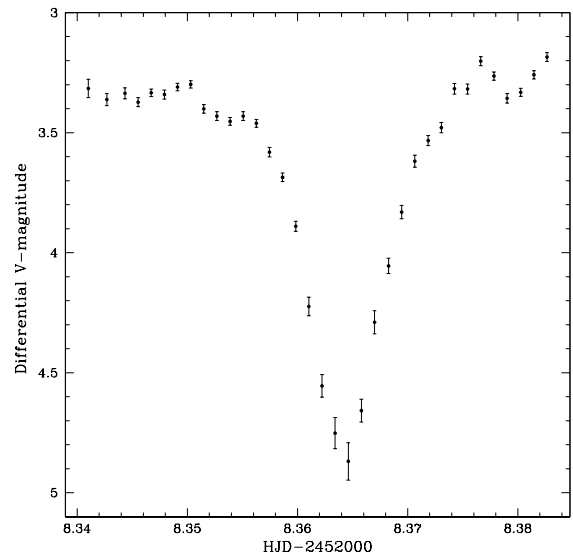,width=8cm}}
  \caption{$V$-band eclipse light curve of V348 Pup.}
  \end{center}
\end{figure*}

\begin{figure*}
\begin{center}
\mbox{\epsfig{file=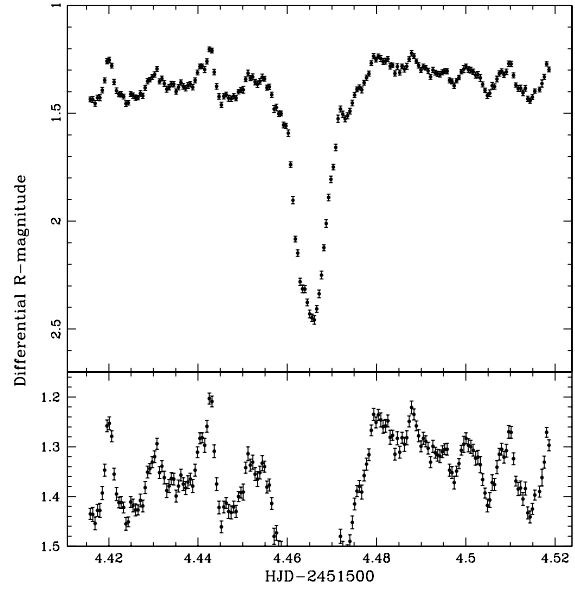,width=8cm}}
\caption{$R$-band light curve of V348 Pup. The presence of pulses is best seen in the lower panel, where the deep eclipse is not plotted.}
\end{center}
\end{figure*}





\begin{figure*}
\begin{center}
\mbox{\epsfig{file=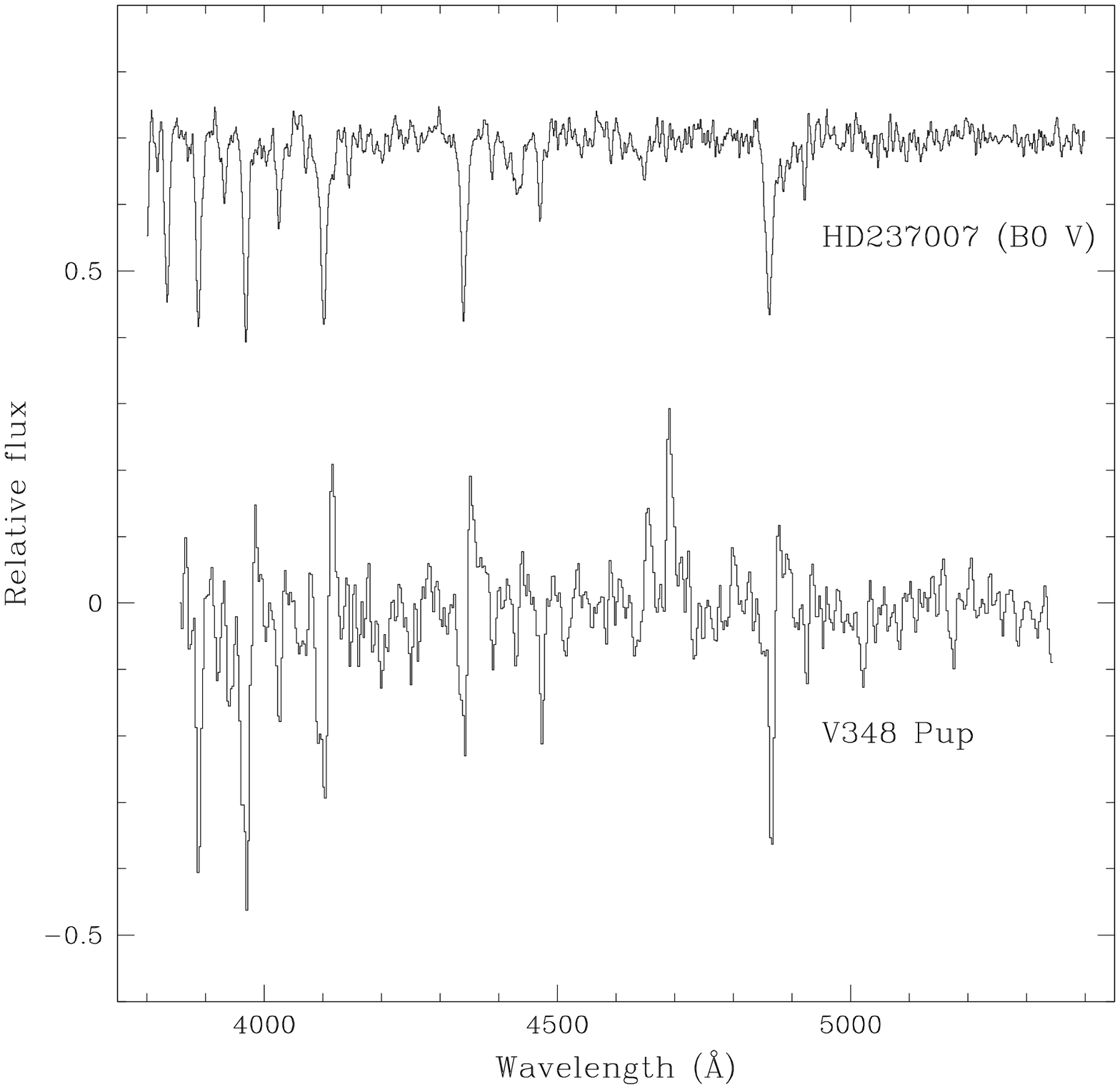,width=8cm}}
\caption{Spectrum of the absorption component in V348 Pup compared to the absorption spectrum of a B0 V star. The stellar spectrum has been continuum-normalized and shifted in flux for clarity.}
\end{center}
\end{figure*}

\label{lastpage}


\begin{thebibliography}{}
\bibitem{baptista96} Baptista R., Patterson J., O'Donoghue D., Buckley D., Jablonski F., Augusteijn T., Dillon W., 1996, IAUC, 6327
\bibitem{casa} Casares J., Mart\'\i nez-Pais I. G., Marsh T. R., Charles P. A., 
L\'azaro C., 1996, MNRAS, 278, 219
\bibitem{v1315dhi} Dhillon V.S., Marsh T.R., Jones D.H.P., 1991, MNRAS, 252, 342
\bibitem{dh1} Dhillon V.S., Jones D.H.P., Marsh, T R., 1994, MNRAS, 266, 859 
\bibitem{groot01} Groot P.J., Rutten R.G.M., van Paradijs J., 2001, A\&A, 368, 183
\bibitem{hellier94} Hellier C., Robinson E. L., 1994, ApJL, 431, 107
\bibitem{hellier96} Hellier C., 1996, ApJ, 471, 949
\bibitem{b4} Horne K., 1986, PASP, 98, 609
\bibitem{h2} Horne K., 1999, in ASP Conf. Ser. 157, Annapolis Workshop on 
Magnetic Cataclysmic Variables, ed. C. Hellier \& K. Mukai (San Francisco: ASP), 349
\bibitem{jacoby84} Jacoby G.H., Hunter D.A., Christian C.A., 1984, ApJS, 56, 257 \bibitem{doppler} Marsh T.R., Horne K., 1988, MNRAS, 235, 269
\bibitem{b4bis} Mart\'\i nez-Pais I.G., Rodr\'\i guez-Gil P., Casares J., 1999, MNRAS, 305,661
\bibitem{eps} Patterson J., 1998, PASP, 110, 1132
\bibitem{sh795} Patterson J., Skillman, D.R., 1994, PASP, 106, 1141
\bibitem{wxari} Rodr\'\i guez-Gil P., Casares J., Dhillon V.S., Mart\'\i nez-Pais I.G., 2000, A\&A, 355, 181 
\bibitem{b5} Rodr\'\i guez-Gil P., Casares J., Mart\'\i nez-Pais I.G., Hakala P., Steeghs D., 2001, ApJL, 548, 49
\bibitem{b6} Rolfe D.J., Haswell C.A., Patterson J., 2000, MNRAS, 317, 759
\bibitem{b7} Rosen S.R., Clayton K.L., Osborne J.P., McGale P.A., 1994, MNRAS, 269, 913
\bibitem{b8} Rutten R.G.M., van Paradijs J., Tinbergen J., 1992, A\&A, 260, 213
\bibitem{m2} Smith D.A., Dhillon V.S., 1998, MNRAS, 301, 767
\bibitem{sdm} Smith D.A., Dhillon V.S., Marsh T.R., 1998, MNRAS, 296, 465 
\bibitem{still} Still M.D., Dhillon V.S., Jones D.H.P., 1995, MNRAS, 273, 863
\bibitem{b10} Szkody P., Pich\'e F., 1990, ApJ, 361, 235
\bibitem{b11} Thomas G.R., 1993, BAAS, 25, 909
\bibitem{b12} Thorstensen J.R., Ringwald F.A., Wade R.A., Schmidt G.D., Norsworthy J.E., 1991, AJ, 102, 272
\bibitem{b13} Tuohy I.R., Remillard R.A., Brissenden R.J.V., Bradt H.V., 1990, ApJ, 359, 204
\bibitem{b14} Warner B., 1995, Cataclysmic Variable Stars, Cambridge University Press, Cambridge



\end{thebibliography}
\end{document}